\documentclass[%
 reprint,
amsmath,
amssymb,
aps,
prl,
floatfix,
superscriptaddress, 
]{revtex4-2}
\usepackage{}
\usepackage{graphicx}
\usepackage{dcolumn}
\usepackage{bm}
\usepackage{siunitx}
\usepackage{hyperref}
\usepackage{braket}
\usepackage[caption=false]{subfig}
\newcommand{\PrYSO}{Pr$^\textup{3+}$:Y$_2$SiO$_5$}
\usepackage[final]{changes}

\makeatletter
\renewcommand{\p@subfigure}{\thefigure}
\makeatother

\begin{document}

\preprint{APS/123-QED}

\title{Quantum storage of qubits in an array of independently controllable solid-state quantum memories}
\author{Markus Teller}
\author{Susana Plascencia}
\author{Samuele Grandi}
\affiliation{ICFO-Institut de Ciencies Fotoniques, The Barcelona Institute of Science and Technology, 08860 Castelldefels (Barcelona), Spain}%
\author{Hugues de Riedmatten}
\affiliation{ICFO-Institut de Ciencies Fotoniques, The Barcelona Institute of Science and Technology, 08860 Castelldefels (Barcelona), Spain}%
\affiliation{ICREA-Institucio Catalana de Recerca i Estudis Avanç ats, 08015 Barcelona, Spain}
\date{\today}

\begin{abstract}
Random-access quantum memories may offer computational advantages for quantum computers and networks. In this paper, we advance arrays of solid-state quantum memories towards their usage as random-access quantum memory. We perform quantum storage of path and time-bin qubits implemented with weak coherent states at the single-photon level, in an array of ten temporally-multiplexed memory cells with controllable addressing. The qubits can be stored in arbitrary combinations of memory cells, from which they are read-out on demand. We find average fidelities of $95_{-2}^{+2}\;\%$ for path qubits and $91^{+2}_{-2}\;\%$ for time-bin qubits. The measured fidelities violate the classical bounds for both encodings and for all ten cells. We also sequentially store a time-bin qubit in two different memory cells, maintain both qubits simultaneously in the array, and perform a collective read-out. The individual control paired with high storage fidelity represents a significant advance towards a solid-state  random-access quantum memory for quantum repeaters and photonic quantum processors.
\end{abstract}

\maketitle


\section{Introduction}
\label{sec:introduction}
In today's classical computation and communication networks, processed information is temporarily stored in its allocated memory. In analogy to a random-access memory for classical computers, a random-access quantum memory (RAQM) connected to quantum processors will perform intermediate storage of quantum information for synchronization between distant quantum processors or repeaters, enabling a wide range of quantum computing and network applications~\cite{Giovannetti2008,Briegel1998,Sangouard2011}. A RAQM should have the capability of storing many qubits in independent modes and provide individual
addressing and on-demand read out of each qubit in a programmable order \cite{Jiang2019}. 
Near-term long-distance quantum communication networks would benefit from RAQMs, as quantum repeater architectures with RAQMs could store many quantum states at the nodes of a repeater link, with high fidelity and efficiency, while waiting for the heralding signal of a successful entanglement with the adjacent node~\cite{Briegel1998,Duan2001,Sangouard2011}. The capacity of the quantum channel connecting the nodes is thus enhanced by the multimodality of the RAQM, which, contrary to single-mode architecture, can store more than one quantum state per communication time \cite{Sangouard2011}. Moreover,  the ability to selectively read out the stored qubits enables additional capabilities for qubit routing and manipulation \cite{Collins2007}.  
Photonic quantum computers would also profit from a RAQM system, as these platforms rely on the successive generation of photons preparing a large photonic cluster state~\cite{Raussendorf2001,Raussendorf2003,Briegel2009}. The memory could buffer the incoming photons during the generation process of the cluster state until the size the quantum resource has reached the desired dimensionality and complexity. A RAQM could also enable the implementation of measurement and feedforward operations on parts of the photonic cluster state while holding the remainder of the quantum resource~\cite{Briegel2009}.

A requirement for the before mentioned applications is the compatibility of the memory with photonic quantum information --  typically encoded in polarization or time degree --  either natively or through conversion. While recently photonic quantum bits (qubits)  have been stored with random-access control in memory arrays of cold-atomic ensembles~\cite{Pu2017, Jiang2019} with over-thousand qubit manipulations~\cite{Zhang2024} and single-atoms coupled to optical cavities~\cite{Langenfeld2020}, these experiments were limited to storage of qubits in the path \cite{Jiang2019} or polarization degree of freedom~\cite{Langenfeld2020}.
Storage of time-bin qubits required additional conversion into the path degree, either realized with hundreds of meters of optical fibers~\cite{Zhang2024} or through optical switching between memory cells~\cite{Zhang2024a}. These conversion processes are resource costly as the additional optical fibers lead to photon loss and optical mapping of a $d$ dimensional quantum state requires $d$ memory cells~\cite{Zhang2024a}.
Additionally, both atomic ensembles and single atoms in optical cavities have intrinsic limits to their scalability in number of cells. For the former, the size of the atomic cloud and the spatially-inhomogeneous atomic density set a transversal limit to the total number of cells with equal performances which can be addressed \cite{Pu2017}.  For the latter, the number of single atoms that can be efficiently coupled to an optical cavity is limited by the size of the cavity mode to tens of atoms~\cite{Hartung2024}.

In this paper, we experimentally explore arrays of rare-earth-doped solid-state quantum memories towards their potential use as RAQMs for path and time-bin quantum information. Ten individually-controlled memory cells are prepared with the atomic-frequency comb (AFC) protocol, whose intrinsic temporal multi-modality allows the storage of time-bin encoded quantum states in each memory cell without the need for conversion~\cite{Afzelius2009,Clausen2011,Gundogan2012}. After absorption from the comb, the application of control pulses (CPs) enable on-demand storage and retrieval\deleted{at the single-photon level}~\cite{Afzelius2010a,Seri2017,Rakonjac2021}. \added{Using weak coherent states (WCS) at the single-photon level, we} \deleted{We} store qubits of two different encodings, time-bin or path, in arbitrary pairs of memory cells and use the individual control over them to tomographically reconstruct the state of the retrieved qubits. We assess their fidelity across the full register and investigate the effect of the detection window on the fidelity as well as on the detection rate. Finally, exploiting the temporal multimodality of each memory cell, we sequentially load two of them with a time-bin qubit each and by leveraging the on-demand operation of the AFC we interfere the two stored qubits with a simultaneous read-out. The array thus acts as an interferometer allowing us to directly probe the coherence between the retrieved states. The solid-state crystal features a homogeneous atomic density across its cross-section and a wide inhomogenous broadening, thus potentially allowing storage over thousands of modes with an uniform efficiency distribution across the array \cite{Ortu2022b}.

\section{Experimental details}
\begin{figure*}[htpb]
    \subfloat{
        \includegraphics[width=1.5\columnwidth]{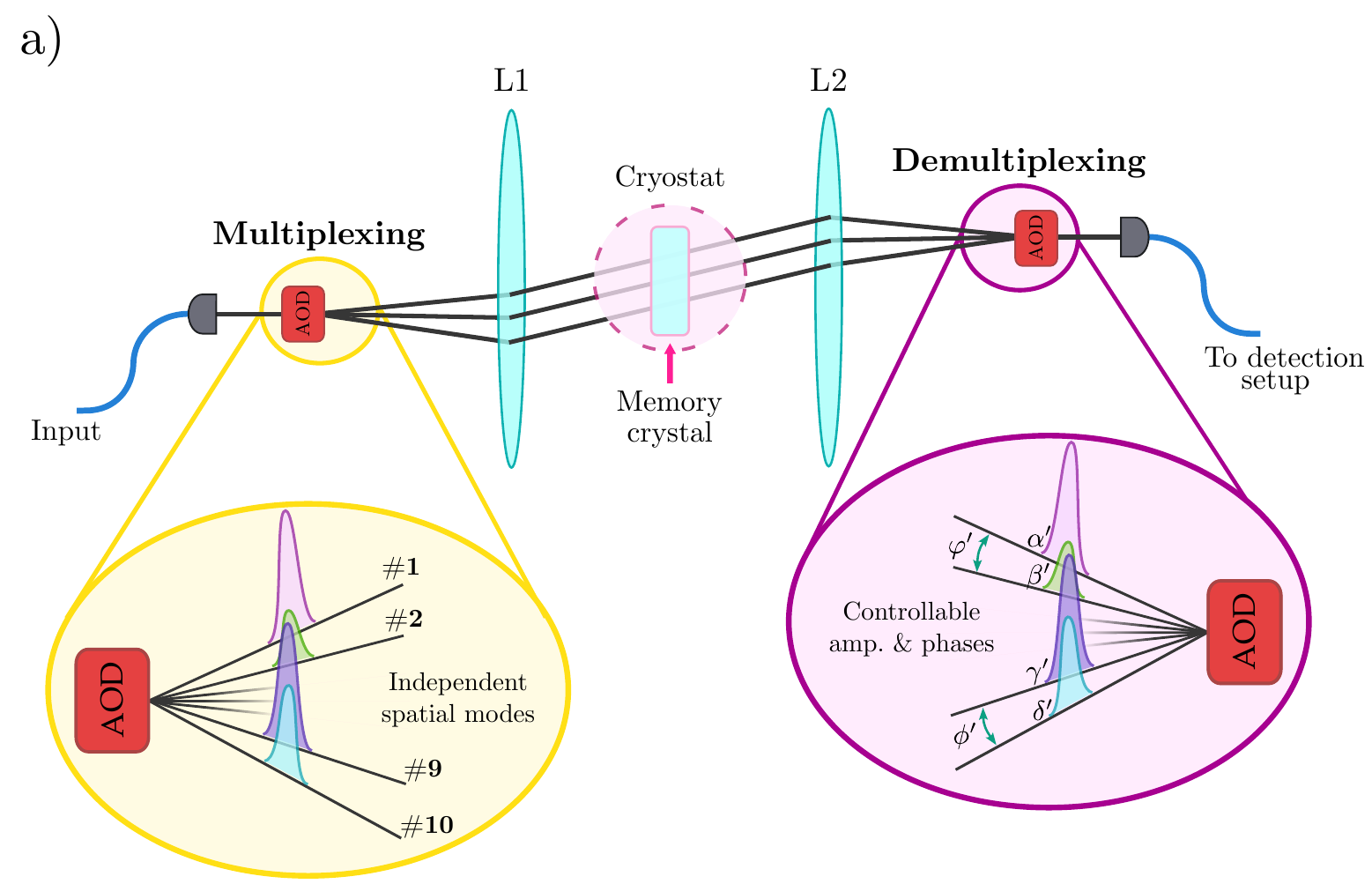}
        \label{fig1:setup_a}}
    \subfloat{
        \includegraphics[width=0.55\columnwidth]{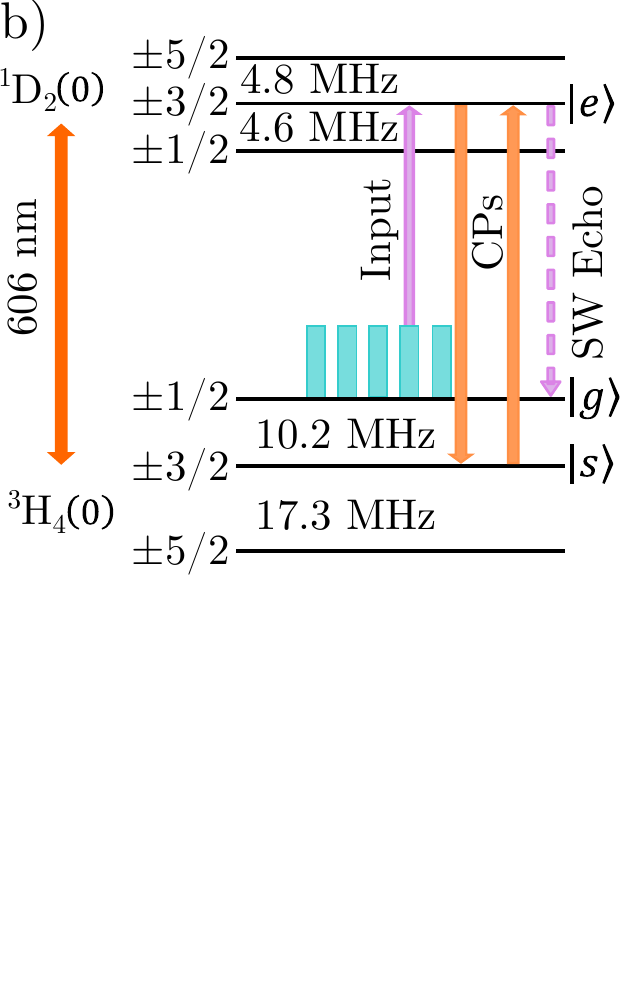}
        \label{fig1:setup_b}}
    \caption{a) Experimental setup of the solid-state quantum memory array. The input light is brought by optical fiber and sent through an acousto-optic deflector (AOD), whose amplitude and phase control allows encoding of arbitrary path qubit states. The light is focused by a lens onto ten independent memory cells in the Pr-doped crystal located in a cryogenic environment. The emitted light is collected by a second lens and sent to a second AOD. Its radio-frequency control projects the qubit states onto a measurement basis and redirects all spatial modes into one. The light is then guided by an outcoupling fiber to a filter and detection setup. b) Structure of the relevant levels of Pr$^{3+}$. The $^{3}H_4(0)$ ground and $^{1}D_2(0)$ excited state are separate by $\SI{606}{\nano\meter}$, each splitting of into three levels due to the spin-orbit coupling. The atomic-frequency comb (AFC) is prepared on the transition from $\ket{g} = \pm 1/2_g$ to $\ket{e} = \pm 3/2_e$. The spin-wave storage is performed by control pulses (CPs) on the $\ket{s} = \pm 3/2_g$ to $\ket{e} = \pm 3/2_e$ transition. After successful storage of an input on $\ket{g}$ to $\ket{e}$, the retrieved spin wave (SW) echo is emitted in the same transition.}\label{fig1:setup}
\end{figure*}

The experimental apparatus is illustrated in Fig.~\ref{fig1:setup_a} and at its heart there is a praseodymium (Pr) doped crystal cooled to a temperature of $\SI{3}{K}$ in a cryogenic environment. The memory array is realized by two acousto-optic deflectors (AODs) and two lenses, symmetrically placed around the crystal. By means of the first AOD, we send light from an optical fiber into up to ten individual memory cells or five memory pairs, accessing spatial multiplexing in our memory crystal. The cells are $\SI{200}{\micro\meter}$ apart and separated by a frequency of $\SI{1}{\mega\hertz}$, which is orders of magnitudes smaller than the inhomogeneous broadening of \PrYSO. For all experiments presented in this paper, we use memory pairs formed by the nearest neighbours of the memory array, where we write a path qubit. By adjusting the amplitudes and phase of the radio-frequency signals applied to the multiplexing AOD we encode an arbitrary qubit state $\ket{\psi} = \alpha\ket{0} + \beta\mathrm{e}^{i\phi}\ket{1}$ in the path degree of freedom using two memory cells $\ket{0}$ and $\ket{1}$\cite{SuppMat}. The second AOD projects the memory output onto an arbitrary measurement basis $\bra{\psi'} = \alpha'\bra{0}+\beta'\mathrm{e}^{i\phi'}\bra{1}$, thus enabling full-state tomography~\cite{James2001}. We emphasize that this implementation with AODs allows us to store in arbitrary combinations of memory pairs including non-neighboring cells. For the storage of time-bin qubits we instead set the amplitudes of the AODs such that we write in single memory cells, but can read in multiple ones at the same time to interfere the qubits stored in different cells. For both path and time-bin encoding the second AOD acts a spatial demultiplexer, erasing the frequency shift imparted by the first AOD and redirecting the output of all the cells into a single spatial mode. The out-coupling fiber then guides the light to a filtering setup after which it is detected. Note that additional optical paths not shown in Fig.~\ref{fig1:setup_a} are used for the memory preparation and application of CPs. A complete description of the experimental setup, including \added{the switching times of the AODs and} the optical paths for the AFC preparation and CPs, is provided in Refs.~\cite{SuppMat, Teller2025a}. 

The level structure of Pr$^{3+}$ relevant for the presented experiments is shown in Fig.~\ref{fig1:setup_b}. The two relevant electronic states are the  $^{3}H_4(0)$ ground and $^{1}D_2(0)$ excited state, separated by a transition at a wavelength of $\SI{606}{\nano\meter}$. Both states split into three sub-levels~\cite{Afzelius2009}: the ground-states $\ket{g} = \pm 1/2_g$, $\ket{s} = \pm 3/2$, and $\ket{aux} = \pm 5/2_g$ are separated by $\SI{10.2}{\mega\hertz}$ and  $\SI{17.3}{\mega\hertz}$; the excited states $1/2_e$, $\ket{e} = \pm 3/2_e$, and $\pm 5/2_e$ are separated by $\SI{4.8}{\mega\hertz}$ and $\SI{4.6}{\mega\hertz}$. The transition from $\ket{g}$ to $\ket{e}$ is where the AFC is prepared, and therefore it is the frequency of the input light. The CPs transfer collective optical atomic excitations in the excited state $\ket{e}$ to collective spin excitations (spin waves) in the ground-state $\ket{s}$, allowing on-demand read-out of the memory. For all experiments, an AFC with $\SI{10}{\micro\second}$ storage time is simultaneously prepared in all the memory cells using an additional AOD line~\cite{Teller2025a}. After successful AFC absorption and a CP, the input light is stored for a variable time. Then, after a second CP, the output is re-emitted on the transition from $\ket{e}$ to $\ket{g}$.

\section{Storage in path encoding}

\begin{figure*}[htpb]
    \centering
    \includegraphics[width=0.8\linewidth]{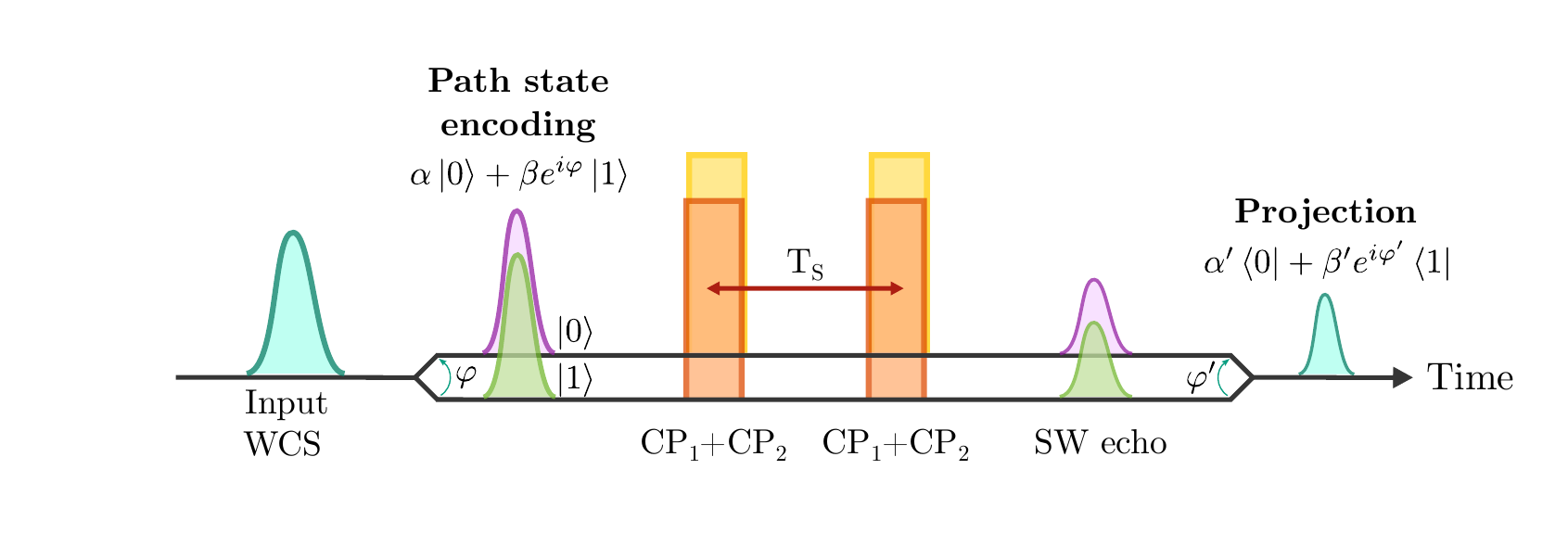}
    \caption{Experimental sequence for storage of path qubits. As first step, the AFC burning and optical pumping steps are performed simultaneously for all ten memory cells. After the memory preparation, the storage cycle starts for the first memory pair: a weak coherent state is sen\replaced{t}{d} as input into the first memory pair forming a path qubit stored in the cells. A $\SI{2.8}{\micro\second}$ long CP is applied to both cells simultaneously to transfer the atomic populations from $\ket{e}$ to $\ket{s}$. After a storage time of $\SI{8}{\micro\second}$ in the spin state, a second CP transfers the atomic populations back to $\ket{e}$. The light will then be emitted after $\SI{5.2}{\micro\second}$. Finally, the storage cycle is repeated for the next memory pair.}
    \label{fig:sequence}
\end{figure*}

We now turn to the storage of path qubits in our solid-state memory array. To assess the fidelity of this process, we store one of the six qubit states  $\ket{\psi}\in\{\ket{0},\ket{1}, \ket{X}=1/\sqrt{2}(\ket{0} + \ket{1}),\ket{-X}=1/\sqrt{2}(\ket{0} - \ket{1}),\ket{Y}=1/\sqrt{2}(\ket{0} + i\ket{1}),\ket{-Y}=1/\sqrt{2}(\ket{0} - i\ket{1}) \}$ in each memory pair and perform full-state tomography~\cite{James2001}.  
The experimental sequence is illustrated in Fig.\ref{fig:sequence}. A Lorentzian-shaped weak coherent pulse with a full-width-at-half-max of $\SI{133(2)}{\nano\second}$ and an average photon number $\bar{n}=1.10(6)$ measured in front of the crystal is sent as input into a memory pair.  The amplitudes and phase of the radio-frequency fields applied to the first AOD are set to encode the qubit state $\ket{\psi}$. Subsequently, a $\SI{2.8}{\micro\second}$-long CP applied to both cells transfers the atomic population to $\ket{s}$. The envelope of the CPs is Gaussian shaped, and the frequency of the pulses is chirped over $\SI{3.4}{\mega\hertz}$. After a storage time of $T_\mathrm{s} = \SI{8}{\micro\second}$ a second control pulse excites the population from $\ket{s}$ into $\ket{e}$, which results in an output signal after a total storage time of $\SI{18}{\micro\second}$. The amplitudes and phase of radio-frequency signal applied to the second AOD are set to correspond to a projective 
measurement in the basis $\bra{\psi'}\in\{\bra{0},\bra{1}, \bra{X},\bra{-X},\bra{Y},\bra{-Y}\}$. The storage sequence then repeats for all other memory pairs. In each cycle of the cryostat, while the preparation step is performed only once, the storage procedure is repeated for each memory pair 51 times.  

\begin{figure}[!htp]
    \centering
    \subfloat{
        \label{fig:2a}
        \includegraphics[width=0.9\columnwidth]{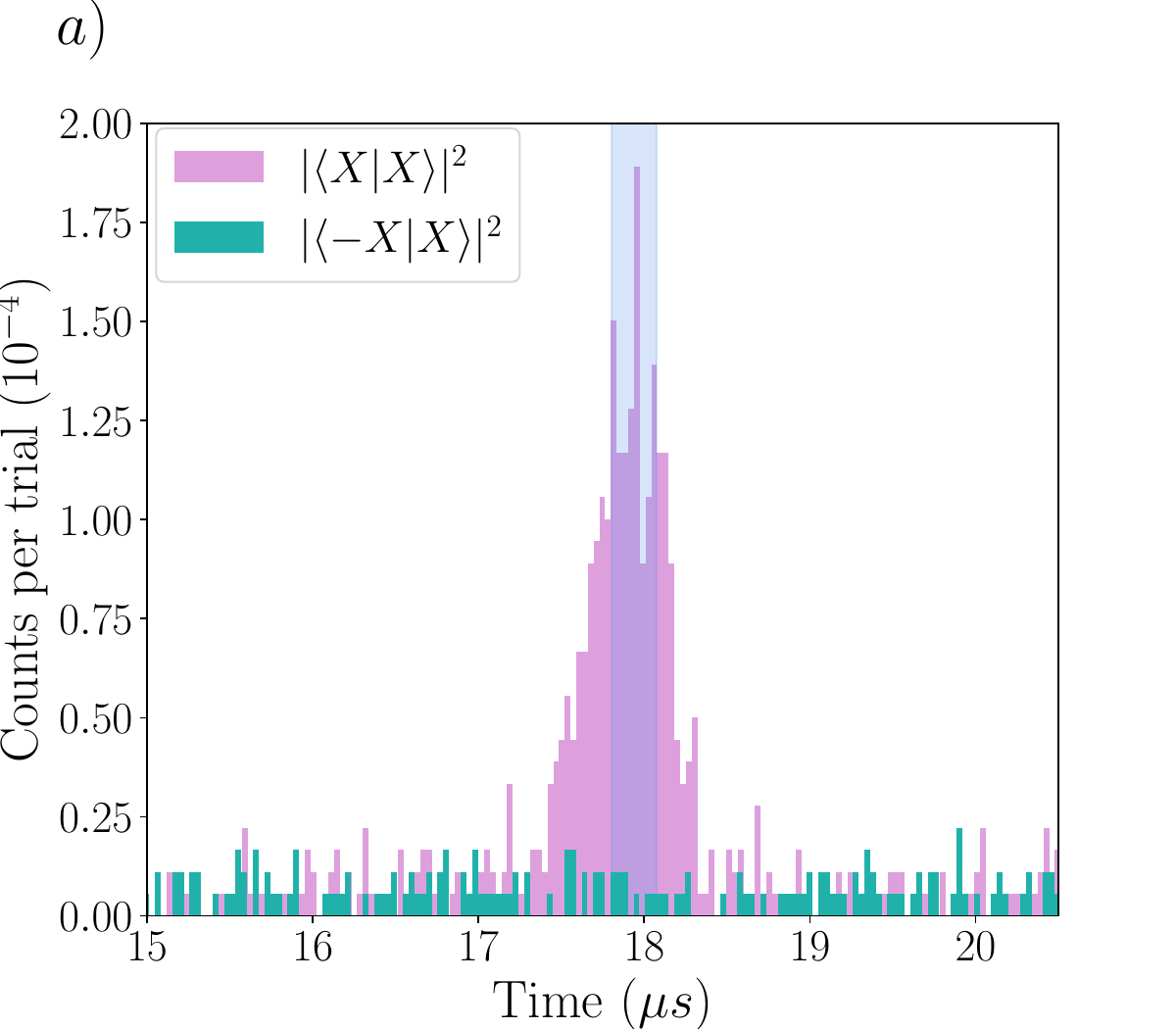}
    }

    \subfloat{
        \includegraphics[width=0.825\columnwidth]{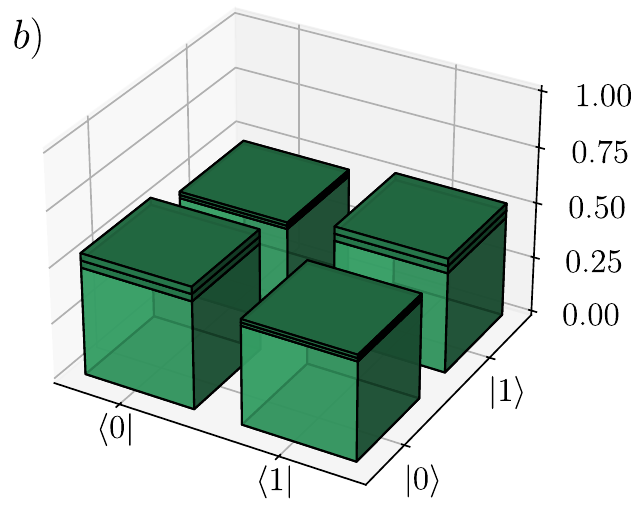}
        \label{fig:2b}
    }


    \caption{Storage and analysis of path qubits. a) Histogram of the detected counts of pair $\{5,6\}$ prepared in $\ket{X}$ and measured in $\bra{X}$ and $\bra{-X}$. The blue area indicates the detection window $t_\mathrm{D} = \SI{270}{\nano\second}$. b) Real part of the reconstructed density matrix $\rho$ for the target qubit state $\ket{X}$ in the memory pair $\{5,6\}$. \added{The uncertainties are represented in dark green.} The imaginary components range from $-0.04$ to $0.04$.} 
\end{figure}

We prepare and measure all six qubit states for all five memory pairs and for all combinations of qubit states and measurement bases. We extract the counts per trial in a detection window $t_\mathrm{D}$ for each combination of qubit state and measurement basis and reconstruct the density matrix~\cite{James2001}. Figure~\ref{fig:2a} shows the counts per trial for the qubit state $\ket{X}$ measured in bases $\bra{X}$ and $\bra{-X}$ for pair $\{5,6\}$ with the detection window $t_\mathrm{D} = \SI{270}{\nano\second}$ indicated as blue area. The input pulse is broadened by the AFC, the CPs and mostly by the filter crystal such that the indicated detection window of $t_\mathrm{D} = \SI{270}{\nano\second}$ contains $42\;\%$ of the detected photon. The real part of the reconstructed density matrix is shown in Fig.~\ref{fig:2b}, and the fidelity of the obtained state $\rho$ with respect to the ideal \added{input} state is $F = |\bra{\psi}\rho\ket{\psi}| = 96^{+1}_{-1}\;\%$. The uncertainties of all presented fidelities are calculated through Monte Carlo resampling considering the Poisson statistics of the detected light~\cite{Efron1993,SuppMat}. The spin-wave memory efficiencies considering the AFC and two CPs range between $1.7(2)$ and $6.7(5)\;\%$ across the array. The efficiencies of the AFC, CPs, and the demultiplexing AOD of each pair are provided in Ref.~\cite{SuppMat}. \added{Note that the filter crystal is prepared with a $\SI{2}{\mega\hertz}$ narrow transparency window, optimized to result in the highest fidelities instead of minimal broadening~\cite{SuppMat}.} 

\begin{figure}[!htp]
    \centering
    \includegraphics[width=0.85\columnwidth]{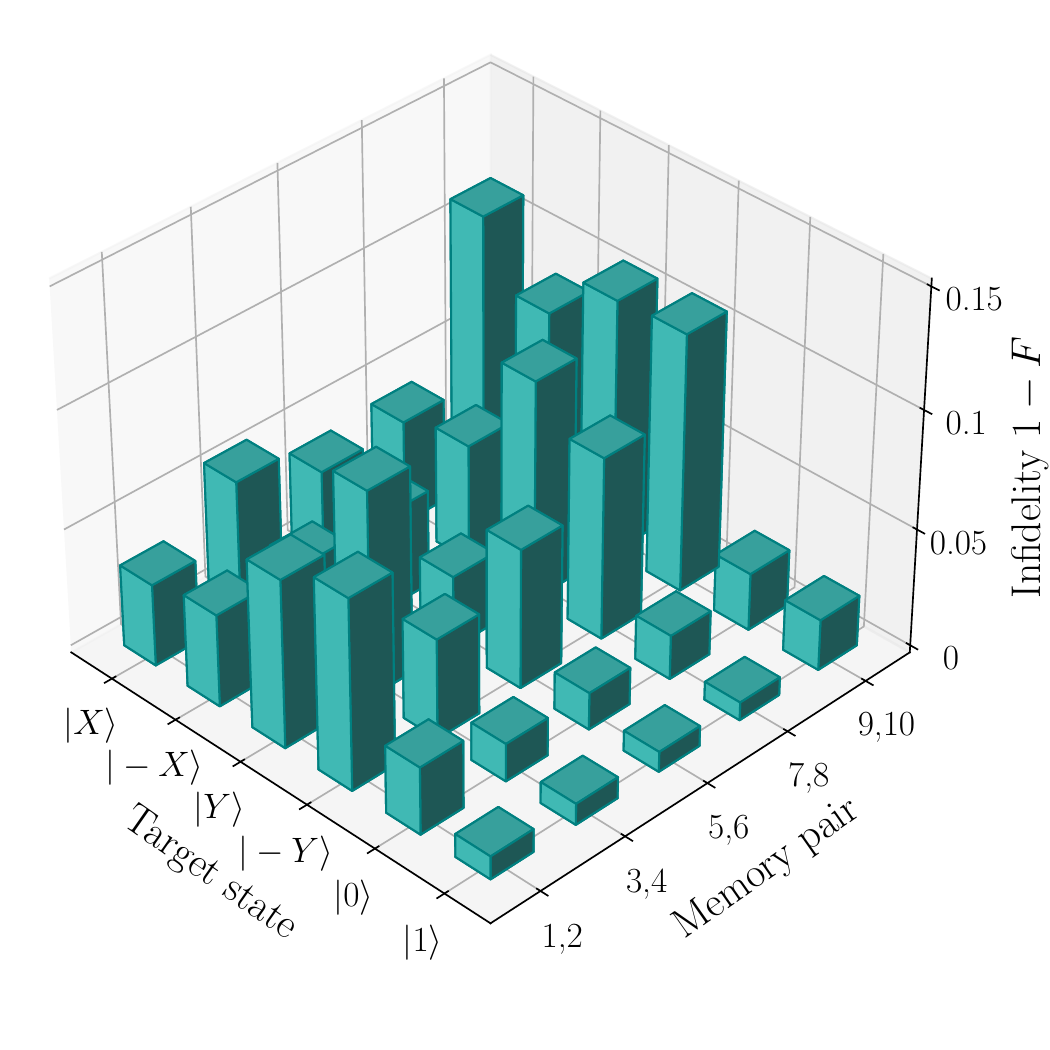}
    
    \caption{Infidelity $1-F$ for each path-encoding target state and memory pair. The density matrix $\rho$ and the infidelities are evaluated for a detection window $t_\mathrm{D} = \SI{270}{\nano\second}$. The data was acquired over $\SI{34}{\hour}$.}\label{fig:2c}
\end{figure}

For all six input qubit states we determine the fidelity and infidelity $1-F$ with respect to the ideal state for $t_\mathrm{D} = \SI{270}{\nano\second}$. The infidelity per qubit basis and memory pair is shown in Fig.~\ref{fig:2c} and varies per target state and memory pair from infidelities as low as $0.8_{-0.3}^{+0.6}\;\%$ for the target state $\ket{1}$ in pair $\{7,8\}$ up to $11_{-4}^{+5}\;\%$ for the target state $\ket{X}$ in pair $\{9,10\}$. The average infidelity across the array and states is $5_{-2}^{+2}\;\%$. Consistent for all pairs, the infidelity of $\ket{1}$ is lower than that of the superposition states $\ket{\pm X}$ and $\ket{\pm Y}$, which we attribute to imbalances in the efficiencies of the individual cells. Prior to the measurements, we calibrate the efficiency for each memory cell measured after the fiber coupling to the detector setup. We adjust the control pulse power for the individual cells such that both cells of each pair have the same efficiency. This balancing ensures that the amplitudes $\alpha$ and $\beta$ of the individual cells are equal to $1/\sqrt{2}$. Nevertheless, over the course of the several tens of hours of measurements, drifts of laser intensities and fiber couplings affect the individual efficiencies and hence imbalance the amplitudes $\alpha$ and $\beta$, affecting most the states $\ket{\pm X}$ and $\ket{\pm Y}$. The individual drifts of the efficiencies of each cell are not enough to alter the signal to noise ratio significantly over the course of the measurements, such that $\ket{0}$ and $\ket{1}$ are mainly limited by fluorescence noise due to the CPs.  Measurements of the efficiencies of the individual cells, a characterization of the noise floor for fluorescence noise and the corresponding signal-to-noise ratio for storage at the single-photon level are presented in Ref.~\cite{Teller2025a}.

\begin{figure}[!htb]
    \centering
    \subfloat{
    \includegraphics[width=1.1\columnwidth]{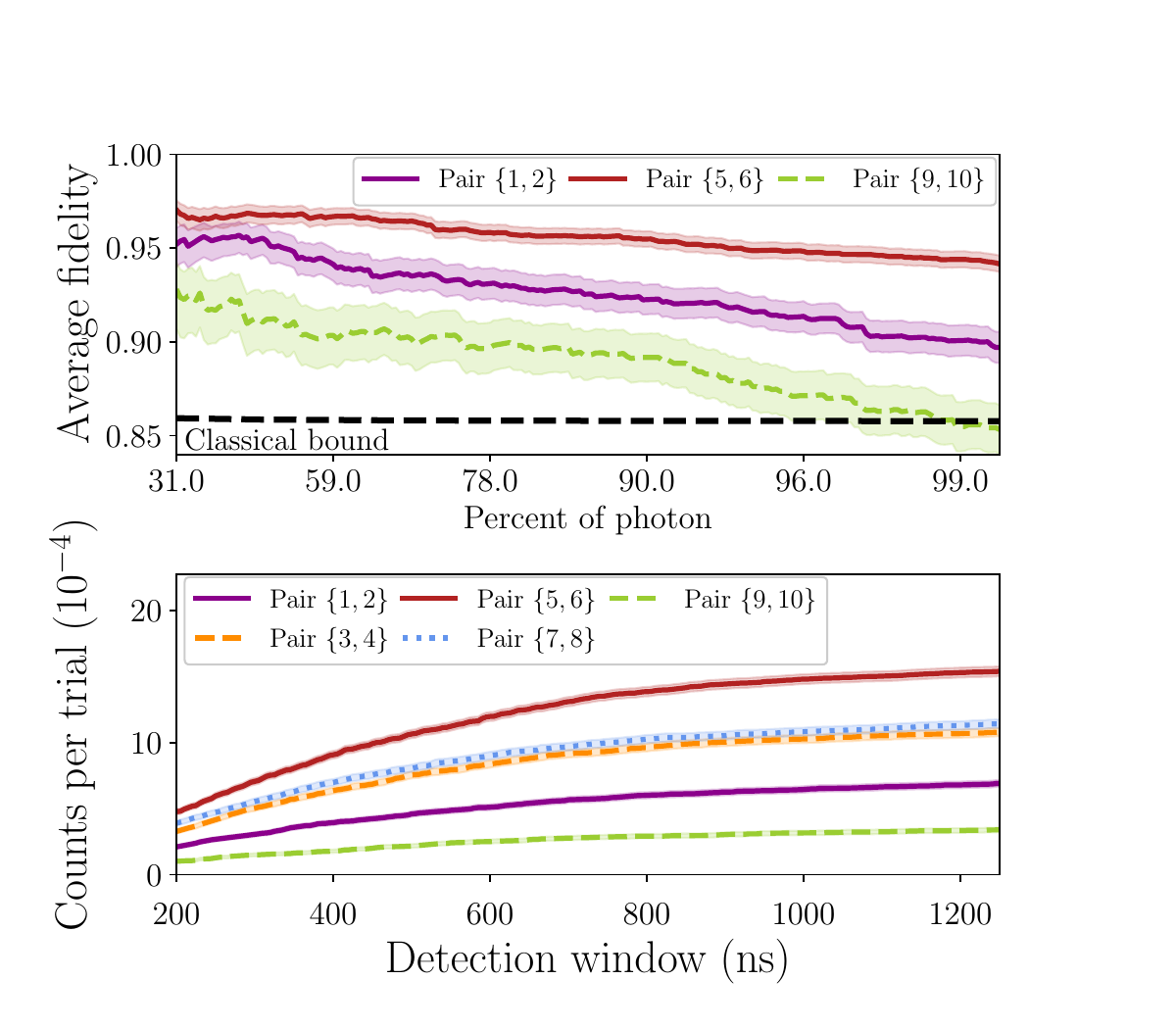} 
    }
    \caption{Top panel: Average fidelity $\bar{F}$ for pairs $\{5,6\}$ and $\{9,10\}$ for varying detection window $t_\mathrm{D}$ for all pairs. \added{The percentage of the photon quantifies how much of the full photon is considered inside a detection window $t_\mathrm{D}$, indicated as shaded area in Fig.~\ref{fig:2a}.} Bottom panel: Detected counts per trial $c_\mathrm{s}$ as function of $t_\mathrm{D}$. The shaded areas indicate the uncertainties.}\label{fig:2d}
\end{figure}

Next, we evaluate the fidelities for detection windows $t_\mathrm{D}$ between $200$ and $\SI{1250}{\nano\second}$ and average the fidelities over all six target states. The average fidelity $\bar{F}$ is shown in Fig.~\ref{fig:2d} for the pair $\{1,2\}$, the pair $\{5,6\}$ with the highest average, and for the pair $\{9,10\}$ with the lowest average across the array. The shaded areas indicate the uncertainty. We compare the average fidelity to a classical bound $F_\mathrm{B}=0.858$, obtained taking into account $\bar{n}$, the efficiencies of the AFC, CPs, and demultiplexing AOD ~\cite{Curty2005,Specht2011,Gundogan2015}. The classical bound takes also into account for the fact that the detection window \added{$t_\mathrm{D}$} only considers a fraction of the detected photon~\cite{SuppMat}. We calculated the violation of the bound $S = (\bar{F}-F_\mathrm{B})/\sigma$ in units of the standard deviation $\sigma$ for detection windows ranging from $200$ to $\SI{1250}{\nano\second}$. We list the the minimum and maximum of $S$ over the range of detection windows in Tab.~\ref{tab:violation_path} together with the value of $S$ considering with $t_\mathrm{D}=\SI{1250}{\nano\second}$ more than $99\;\%$ of the detected photon. Although the average fidelity decreases with increasing $t_\mathrm{D}$, $\bar{F}$ violates $F_\mathrm{B}$ for at least $S = 3.8\sigma$ for all pairs and reaches violations of up to $S = 25.3\sigma$ for pair $\{5,6\}$. The violation decreases from the central pair $\{5,6\}$ of the memory array to the pairs on the sides, which is due to lower signal to noise ratios for pairs at the edges of the array.

To illustrate this dependency on the signal to noise ratio, we plot the counts per storage trial $c_\mathrm{s}$ for varying $t_\mathrm{D}$ for all pairs in Fig.~\ref{fig:2d}. The values for $c_\mathrm{s}$ decrease from the center of the array to the outermost modes from $c_\mathrm{s} = 6.4(2)\cdot10^{-4}$ to $c_\mathrm{s} = 1.4(1)\cdot10^{-4}$ for $t_\mathrm{D}=\SI{270}{\nano\second}$, as expected from the varying efficiencies across the pairs~\cite{SuppMat}. This decrease coincides with the decrease in the average fidelity, as we observe the highest value for $\bar{F}$ for the pair $\{5,6\}$ and the lowest value for the pair $\{9,10\}$. Nevertheless, we emphasize that the counts per storage trial increase by more than factor $2.6$ from $200$ to $\SI{865}{\nano\second}$ while still violating the classical bound by at least $2\sigma$. We thus conclude successful quantum storage of path qubits in five different memory pairs over detection windows up to $\SI{865}{\nano\second}$ containing $87\;\%$ of the detected photon. We highlight that four of five qubits violate the classical bound while considering more than $99\;\%$ of the detected photon.

\begin{table}[htb]
    \centering
    \begin{tabular}{c|c|c|c}
    \hline
        {\bfseries Pair} & {\bfseries Maximum $S$ ($\mathbf{\sigma}$)}  & {\bfseries Minimum $S$($\mathbf{\sigma}$)} & {\bfseries $S$ at 1250 ns ($\mathbf{\sigma}$)} \\
        \hline
         $\{1,2\}$& 10.8 & 4.5& 4.5\\
        $\{3,4\}$ & 15.4 & 10.8&10.8\\
        $\{5,6\}$ & 25.3 & 17.7&18.8\\
       $\{7,8\}$  & 14.9 & 9.4& 9.5\\
       $\{9,10\}$  & 3.8 & -0.5&-0.5\\
    \end{tabular}
    \caption{Maximum and minimum violation of the classical bound in units of standard deviation ($\sigma$) of each memory pair. For comparison, the violation at $t_\mathrm{D} = \SI{1250}{\nano\second}$ is provided.}
    \label{tab:violation_path}
\end{table}

\section{Storage in time-bin encoding}
\begin{figure*}[htpb]
    \includegraphics[width=1\textwidth]{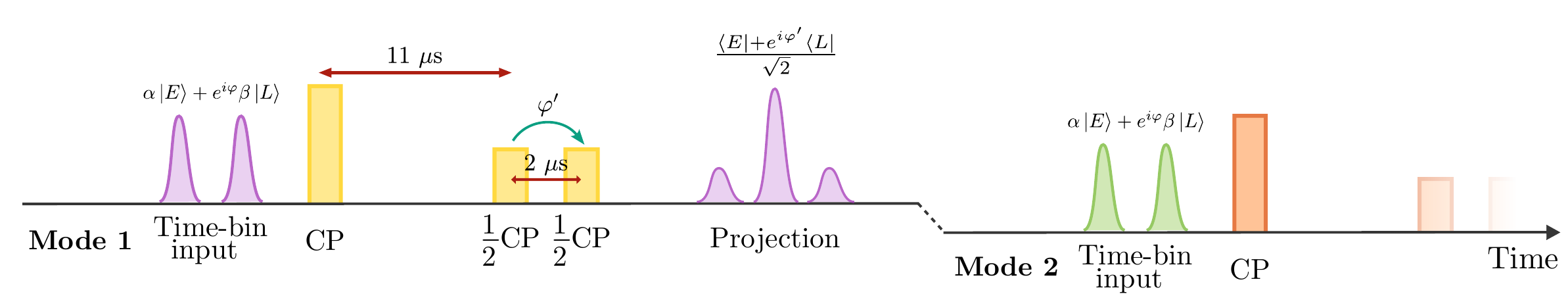}
    \caption{Experimental sequence for storage and analysis of time-bin qubits. After the preparation of all ten memory cells, a time-bin qubit is sent as input into the first cell, where it is stored as a spin wave with the application of a CP. After $\SI{11}{\micro\second}$ of storage in the spin state, a pair of CPs separated by $\SI{2}{\micro\second}$, matching the separation of the time bins, implement a Franson-type interferometer. The time-bin qubit is re-emitted, and the detection signal exhibits three peaks with the central one corresponding to a measurement in the desired basis. The sequence continues with the next memory cell.}
    \label{fig:seq_timebin}
\end{figure*}

Next, we discuss the experiments on the storage of time-bin qubits for which our memory array is particularly well-suited due to its time multiplexing capabilities enabled by the AFC protocol. Without this multiplexing capability, the time bin qubit has to be first converted to a path qubit and then stored in two memory cells, reducing the total capacity of the memory array. Similar to the experiments with path qubits, we prepare one of the six qubit states $\ket{\psi}\in\{\ket{E},\ket{L}, \ket{X},\ket{-X},\ket{Y},\ket{-Y}\}$  in the time-degree of freedom and store the qubits in one of the ten cells of the memory array. Afterwards, we again perform full-state tomography and assess the fidelity of the retrieved states with respect to the ideal case.

\begin{figure}[!htb]
    \centering
    \subfloat{
    \includegraphics[width=1.\columnwidth]{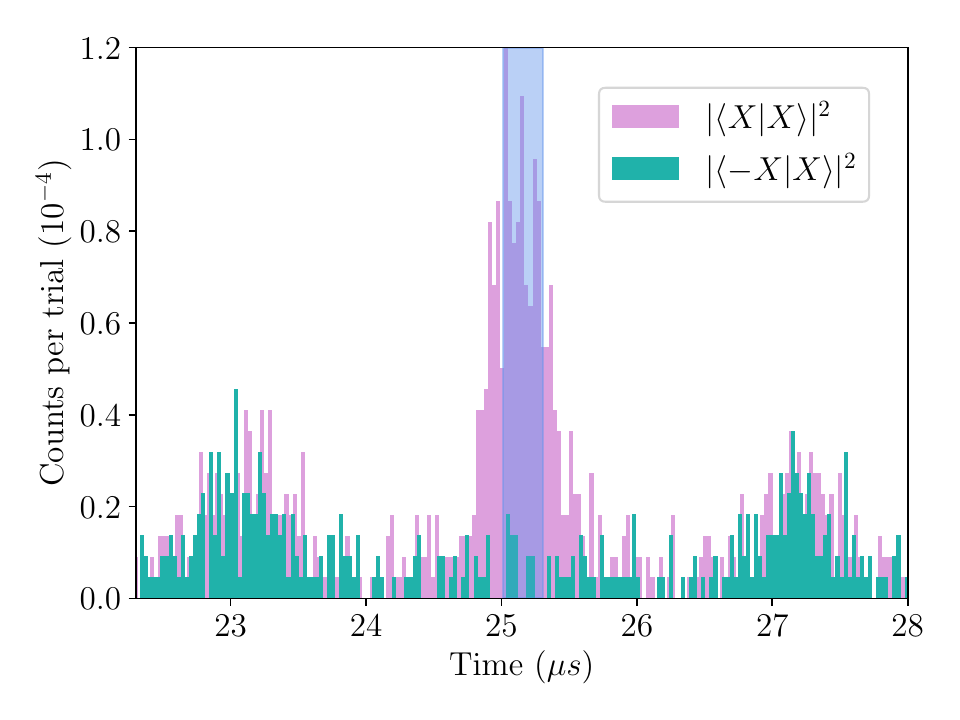} 
    }
    \caption{Histogram of a detected time-bin qubit prepared in $\ket{X}$ and stored in  cell five  and measured in $\bra{X}$ and $\bra{-X}$. The blue area indicates the detection window $t_\mathrm{D} = \SI{295}{\nano\second}$.}\label{fig:3Inter}
\end{figure}

The experimental sequence is illustrated in Fig.~\ref{fig:seq_timebin}. A time-bin qubit is encoded in two Lorentzian-shaped pulses with full-width-at-half-max of $\SI{133(2)}{\nano\second}$ separated by $\SI{2}{\micro\second}$ and a mean photon number of $\bar{n} = 0.94(3)$ measured after the first AOD and integrated over both time bins. The time-bin qubit is then sent to a memory cell, where it is stored as a spin-wave for $\SI{11}{\micro\second}$. To analyse the fidelity of the retrieved time-bin qubit, we perform measurements in the basis $\bra{E}$ or $\bra{L}$ with a single CP that transfers all the population from $\ket{s}$ to $\ket{e}$. We implement measurements in the superposition basis $\bra{\psi'} = 1\sqrt{2}(\bra{E}+\mathrm{e}^{i\Delta \varphi}\bra{L})$ with a solid-state equivalent Franson interferometer~\cite{Franson1989, Timoney2013, Gundogan2013}: we apply two CPs with separation matching the $\SI{2}{\micro\second}$ separation of the time-bin qubit and a relative phase $\Delta \varphi$, selecting one of the measurement basis of $\{\bra{X},\bra{-X},\bra{Y},\bra{-Y}\}$. Each CP transfers half of the atomic population back from $\ket{s}$ to $\ket{e}$, with the early and late control pulse acting as the early and late arms of the interferometer. The detection signal then exhibits three distinct peaks with the central peak corresponding to a measurement in the desired basis, set by the relative phase between the two CPs. Thus, the two half CPs effectively implement a Franson measurement scheme~\cite{Franson1989}. However, we emphasize that in contrast to its implementation with a fiber interferometer, which would require a fiber delay line of $\SI{400}{\meter}$, no active length stabilization of the interferometer is necessary, thus decreasing the experimental complexity. We repeat this storage and analysis sequence for all memory cells cycling through them $41$ times per cycle of the cryostat. As for the experiments with path qubits, the preparation of the memory array is performed only once per cryostat cycle. An example of a detected interference pattern is shown in Fig.~\ref{fig:3Inter} for a time-bin qubit in $\ket{X}$ stored in cell five and measured in $\bra{X}$ and $\bra{-X}$.    

\begin{figure} [htb]
    \subfloat{
        \includegraphics[width=0.75\columnwidth]{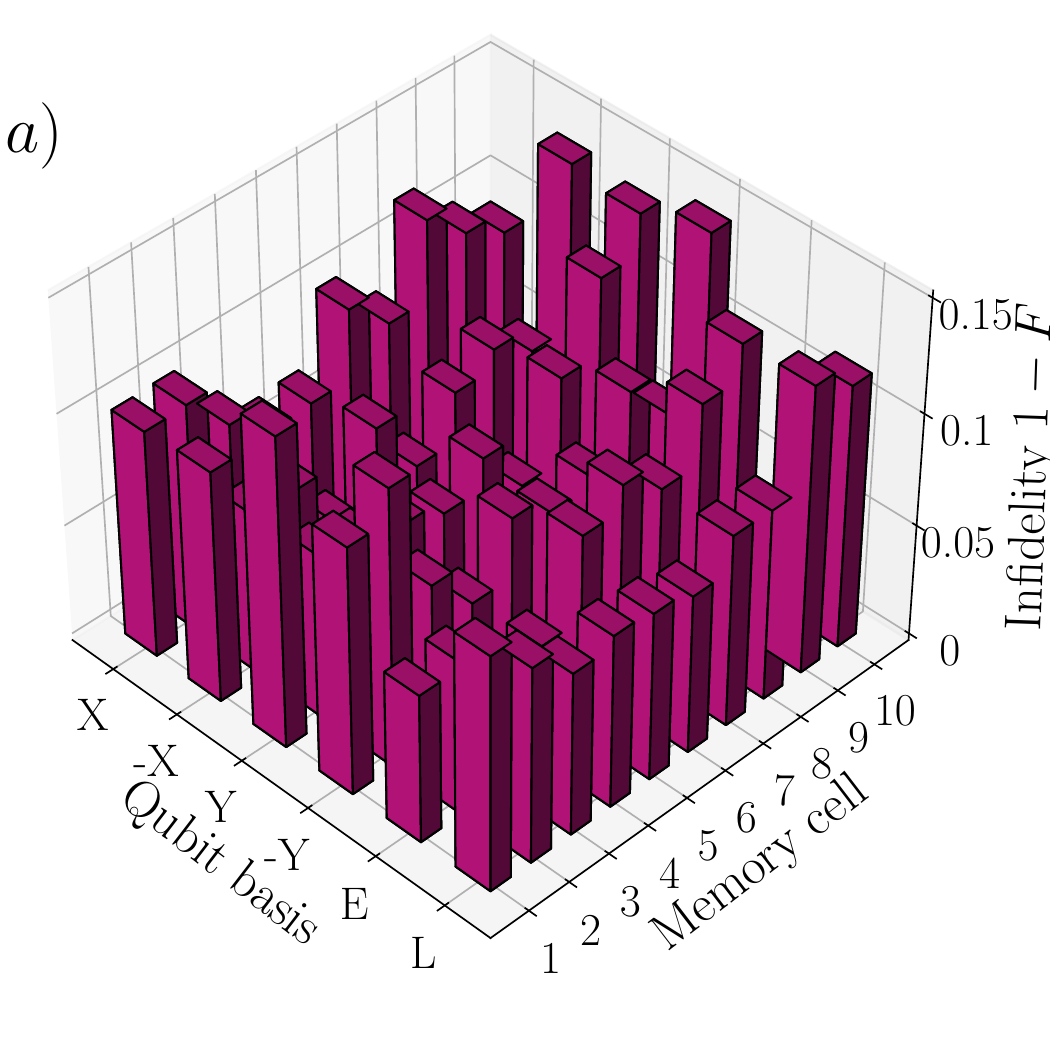}
        \label{fig:time_bin_inf}
    }

    \subfloat{\label{fig:avg_fid_time}
        \includegraphics[width=0.75\columnwidth]{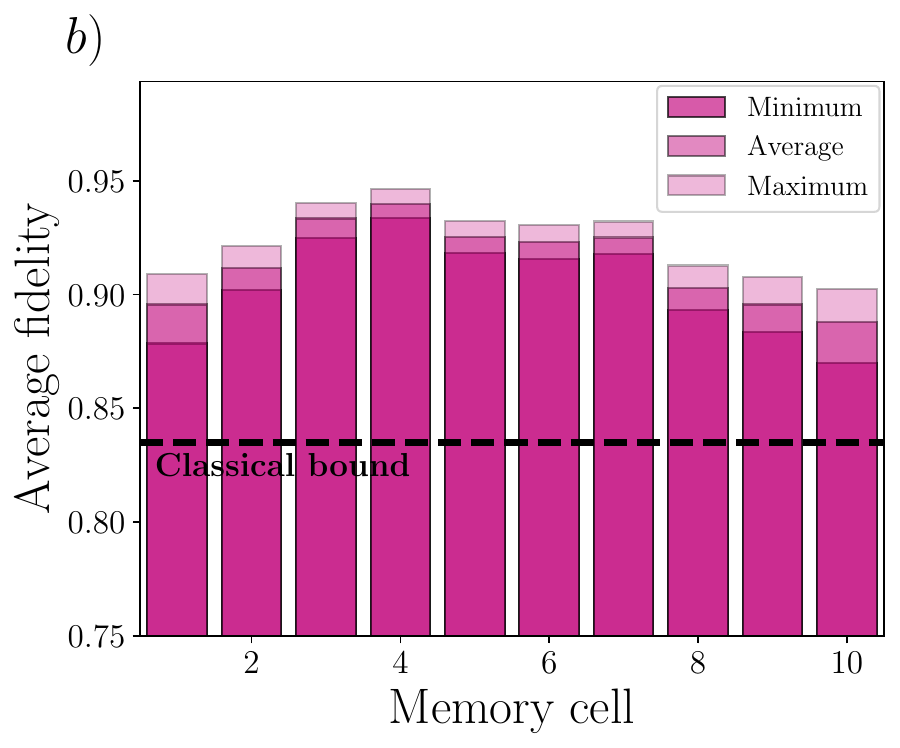}    
    }
    \caption{Storage of time-bin qubits. a) Infidelity $1-F$ per input state and memory cell for a detection window $t=\SI{295}{\nano\second}$. The data was acquired over $\SI{41}{\hour}$ b) Averaged fidelity $\bar{F}$ determined for the fidelities in b). A classical bound $F_\mathrm{B} = 0.835$ is indicates the threshold for quantum storage.} \label{fig:timebin_fid}
\end{figure}

In order to fully gauge the response of our memory array we proceed to store all six input qubits in the time degree of freedom in all memory cells and perform full state tomography to calculate values of fidelity $F$ and infidelity $1-F$ per memory cell with respect to the ideal input states. The infidelities are shown in Fig.~\ref{fig:time_bin_inf} for all combinations of cells and input states considering $t_\mathrm{D} = \SI{295}{\nano\second}$. The infidelities range from $15^{+4}_{-4}\;\%$ to $3^{+3}_{-1}\;\%$ with an average of $9^{+2}_{-2}\;\%$  across the ten cells with decreasing infidelities for cells located more at the center of the crystal. To compare the fidelities to a classical bound, the average fidelity $\bar{F}$ is calculated over the six qubit states, plotted in Fig.~\ref{fig:avg_fid_time}. The classical bound $F_\mathrm{B}  = 0.835$, indicated as dashed line, again considers $\bar{n}$, the efficiencies of the AFC, two-way transfer, the second AOD and the fact that we only consider a fraction of the detected photon~\cite{Curty2005,Specht2011,Gundogan2012, SuppMat}. The efficiencies considered in the bound $F_\mathrm{B}$ are listed in Ref.~\cite{SuppMat}. For $t_\mathrm{D}=\SI{295}{\nano\second}$, the violations of the classical bound $S = (\bar{F}-F_\mathrm{B})/\sigma$ range between $S = 2.9\sigma$ and $16.2\sigma$, indicating quantum storage for all ten cells. 

\begin{figure} [htb]
    \subfloat{
        \includegraphics[width=1.1\columnwidth]{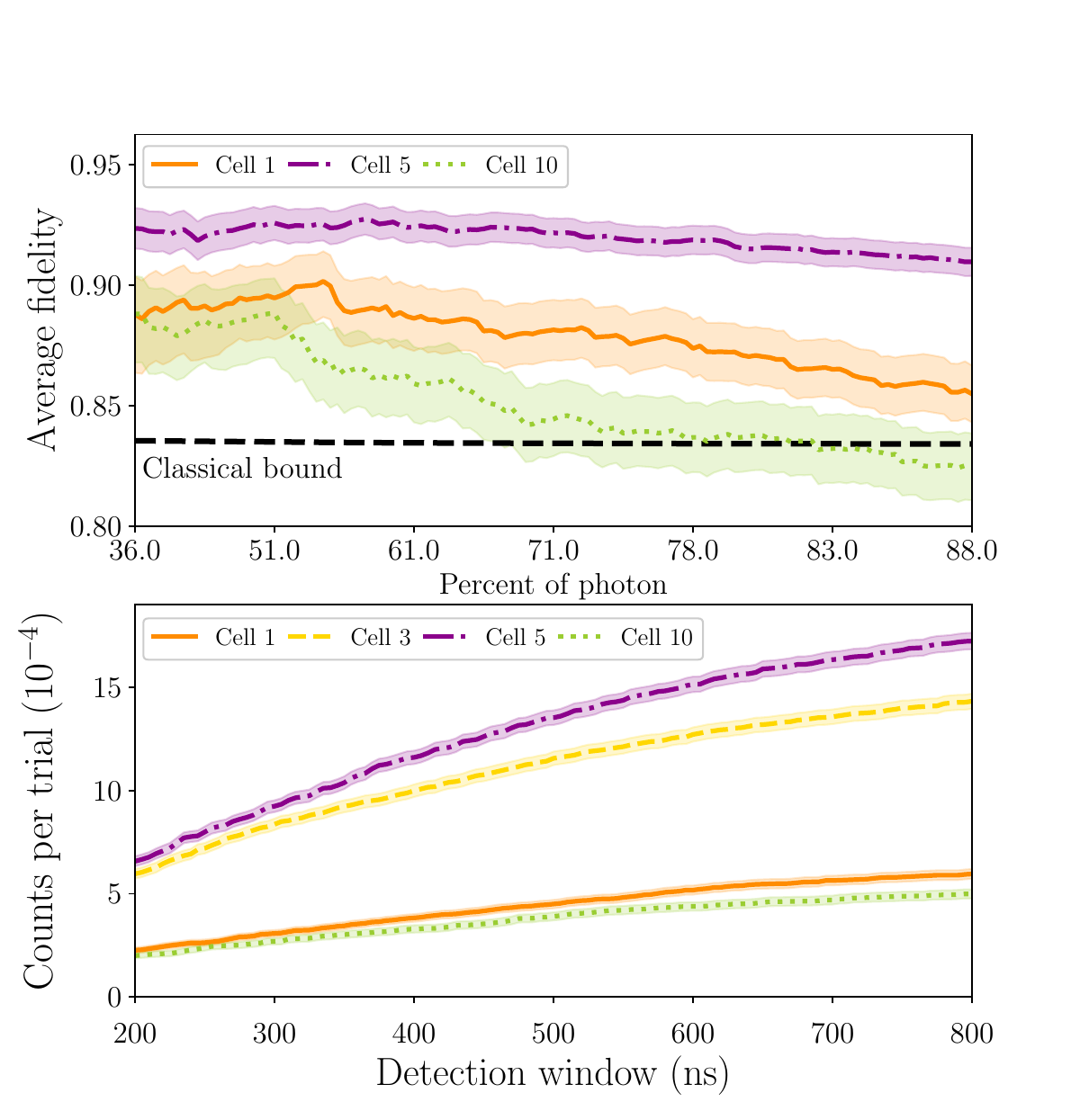}
        \label{fig:tb_fid_window}
    }

    \caption{Average fidelity for cells $1,5,10$  and counts per trial for cells $1,3,5,10$. The classical bound $F_\mathrm{B} = 0.835$ is indicated as horizontal dashed line. The shaded areas indicate the uncertainties. \added{The percentage of the photon quantifies how much of the full photon is considered inside a detection window $t_\mathrm{D}$, indicated as shaded area in Fig.~\ref{fig:3Inter}.}}
    \label{fig:tb_window}
\end{figure}

Next, we evaluate the average fidelity $\bar{F}$ and the violation of the classical bound for detection windows $t_\mathrm{D}$ between $200$ and $\SI{1250}{\nano\second}$. Figure~\ref{fig:tb_window} shows $\bar{F}$ and the counts per trial $c_\mathrm{s}$ for some cells including the ones with the highest and lowest values for $\bar{F}$. Over the range of $t_\mathrm{D}$, the average fidelity decreases while $c_\mathrm{s}$ increases by at least a factor of $2.7$ for all cells. All ten cells violate the classical bound for at least $2\sigma$ up to $t=\SI{335}{\nano\second}$, at which twice the standard deviation of $\bar{F}$ of cell ten intersects with $F_\mathrm{B}$ and  $55\;\%$ of the detected light was considered. The minimum and maximum of $S$ determined for detection windows between $200$ and $\SI{1250}{\nano\second}$ are listed for each cell in Tab.~\ref{tab:violation_tb}. The maximum of $S$ range from $2.9$ for cell ten to up to violations of $19.3$ of cell four. We therefore conclude quantum storage of time-bin qubits in all ten memory cells of our solid-state array. We emphasize that for eight of the ten cells, $S$ is greater than $2\sigma$ for all values of $t_\mathrm{D}$, hence considering the full size of the detected photon.

\begin{table}[htb]
    \centering
    \begin{tabular}{c|c|c|c}
    \hline
        {\bfseries Cell} & {\bfseries Maximum $S$  ($\mathbf{\sigma}$)}  & {\bfseries Minimum $S$ ($\mathbf{\sigma}$)} & {\bfseries $S$ at 1250 ns ($\mathbf{\sigma}$)}  \\
        \hline
         $1$& 4.7 & -1.0 & -1.0\\
         
        $2$ & 8.9 & 3.6 & 3.7\\
        $3$ & 13.3 & 7.6 & 7.8\\
       $4$  & 19.3 & 13.7 & 15.2\\
       $5$  & 14.7 & 10.3 & 11.1\\
       $6$& 14.5 & 10.7 & 11.6\\
        $7$ & 13.8 & 9.4 & 9.9\\
        $8$ & 8.1 & 5.0 & 5.0\\
       $9$  & 5.6 & 2.4 & 2.4\\
       $10$  & 2.9 & -3.0 & -3.0\\
    \end{tabular}
    \caption{Minimum and maximum violation of the classical bound in units of standard deviation ($\sigma$) of each memory cell for detection windows from $200$ to $\SI{1250}{\nano\second}$.}
    \label{tab:violation_tb}
\end{table}

\section{Interference of time-bin qubits}
Finally, we take advantage of the multimodality of our quantum memory array, combining the storage capabilities in path and time degrees to demonstrate simultaneous storage of two time-bin qubits, each in a separate cell of a memory pair. We then probe the coherence between the two qubits and determine the visibility $V$ to show that our memory array can not only store arbitrary qubits but also interfere them pairwise, an important feature for RAQM-like operation.

\begin{figure*} [htpb]
    \includegraphics[width=1\textwidth]{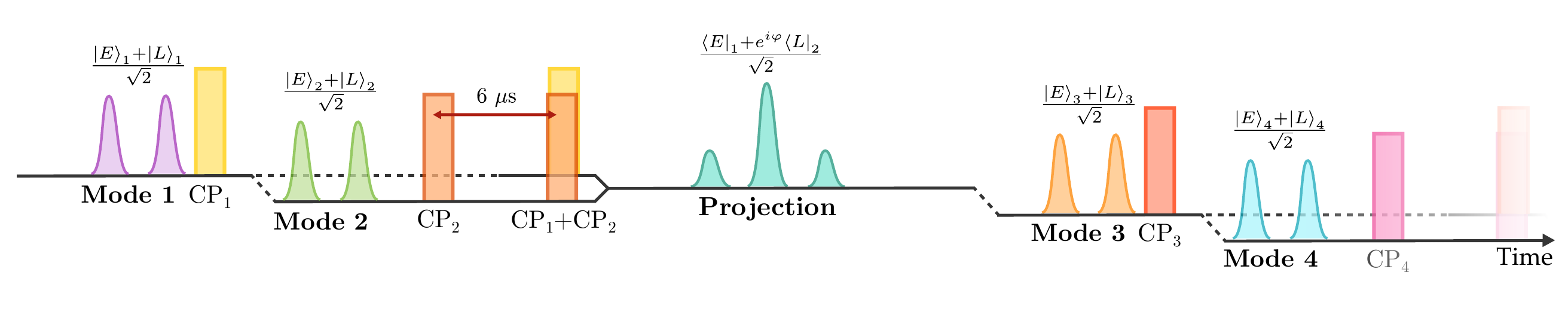}
    \caption{Experimental sequence for simultaneous storage of two time-bin qubits. After preparation of the AFC in the array, the time-bin qubit $\ket{X}$ is sent to a first cell, where it is stored as a spin-wave. A second time-bin qubit in $\ket{X}$ is sent afterwards to a second cell, where the CP for spin-wave storage has an extra phase difference $\Delta \varphi$ and is also sent with an additional $\SI{2}{\micro\second}$ delay with respect to the input. After $\SI{6}{\micro\second}$, a CP acting on both cells recalls the stored qubits simultaneously. The two qubits interfere, and due to the $\SI{2}{\micro\second}$ shift of the second CP the detected interference patter exhibits three peaks, where the amplitude of the middle peak varies with $\Delta \varphi $. The sequence then continues with the next pair of memory cells.}
    \label{fig:seq_interference}
\end{figure*}

The sequence is illustrated in Fig.~\ref{fig:seq_interference}. An input time-bin qubit $\ket{X}$ with $\bar{n} = 0.94(3)$ is sent into the first cell of a memory pair, where it is stored as a spin-wave. 
Then, a second time-bin qubit in $\ket{X}$ is sent to the second cell of the pair, followed by a CP. After $\SI{6}{\micro\second}$, a CP on both cells of the pair transfers the atomic population back to the excited state, leading to a re-emission of both qubits, which are mixed by the second AOD where they interfere. To analyse the coherence between the qubits, we implement a Franson-type interferometer in a similar fashion as in Fig.~\ref{fig:seq_timebin}: We shift the phase of the first CP of the second cell to add a varying phase $\Delta\varphi$ between the re-emitted qubits~\cite{Franson1989} and introduce an extra $\SI{2}{\micro\second}$ waiting time between the input and CP of the second cell. Therefore, the re-emission of the time-bin qubit from the second cell occurs $\SI{2}{\micro\second}$ earlier, forming a detection signal with three distinct peaks. The central peak of this pattern corresponds to the interference between $\ket{E}$ of the first cell and $\ket{L}$ of the second cell. Finally, we proceed to the subsequent memory pair until we concluded all pairs of the array. The steps from storage to recall are repeated $41$ times for each pair.

\begin{figure} [!htb]
    \includegraphics[width=0.9\linewidth]{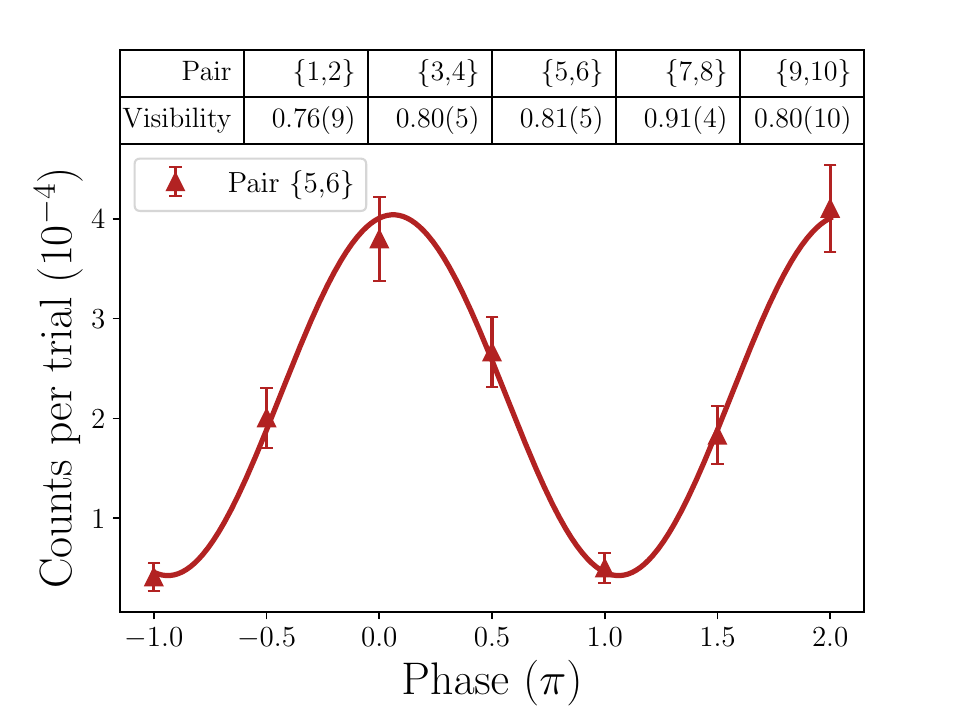}
    \caption{Counts per trial $c_\mathrm{s}$ as function of the analysis phase $\Delta\varphi$ of first CP acting on the second memory cell of pair $\{5,6\}$. The solid line is a fit to $c_\mathrm{s} = A(1+V\cos(\Delta\varphi+\varphi_0))$ with amplitude $A$, phase offset $\varphi_0$. The extracted visibilities for all pairs are listed. The data was acquired over $\SI{8}{\hour}$.}
    \label{fig:fringe}
\end{figure}

We measure the counts per trial $c_\mathrm{s}$ of the interference peak for phases $\Delta\varphi$ ranging from $-\pi$ to $2\pi$ considering a detection window $t_\mathrm{D}=\SI{300}{\nano\second}$ and plot the result for the pair $\{5,6\}$ in Fig.~\ref{fig:fringe}. The detection signals exhibit periodic oscillations with $\Delta \varphi$, to which we fit the function $c_\mathrm{s} = A(1+V\cos(\Delta\varphi+\varphi_0))$. Thus, we extract the visibility $V$ as figure of merit for the coherence between the qubits. The parameters $A$ corresponds to the maximal detections per trial and $\varphi_0$ to an arbitrary phase offset. The extracted visibilities are listed in Fig.~\ref{fig:fringe} with the fit curve of pair $\{5,6\}$ indicated as solid line. The visibilities range between $0.76(9)$ and $0.91(4)$ demonstrating a high-degree of coherence between the qubit pairs, proving the reliability of our multimode memory array. From our extended analysis in Ref.~\cite{SuppMat}, we find that the visibilities are limited by the fluorescence noise from the CPs.

\section{Discussion and Outlook}
In summary, we have demonstrated the storage of path and time-bin qubits in an array of independently controllable  solid-state quantum memories with temporal multiplexing. Five path qubits have been stored and their fidelities analysed by means of full-state tomography using AODs to encode the qubits and to project onto a measurement basis. The violations of the classical bound ranged between $25.3\sigma$ and $3.8\sigma$ indicating quantum storage for all pairs. Using the time-bin encoding, qubits have been stored in ten cells of the memory array. The full-state tomography was implemented in an interferometric fashion by means of the CPs. The fidelities of the reconstructed states violate the classical bound by more than two standard deviations for detection windows up to $\SI{335}{\nano\second}$ for all ten cells. As a step towards the implementation of a RAQM with this solid-state array, we have sequentially loaded and simultaneously stored two time-bin qubits. We have probed the coherence between the stored qubits with interferometric measurements using the visibility $V$ as figure of merit for the coherence. We have determined $V$ to range between $0.76(9)$ and $0.91(4)$, indicating a high coherence between the retrieved qubit states.

As assessed in detail in Ref.~\cite{SuppMat}, the presented storage fidelities and visibilities are limited by the signal-to-noise ratios. With the main noise contribution being the fluorescence induced the CPs, a path to increase the fidelities could be to employ stronger filtering techniques such as using a filtering crystal in double-pass configuration~\cite{Yang2018} or improving the preparation of the individual memories, to bring them to the levels reported in standard bulk memories~\cite{Rakonjac2021}. \added{We emphasize that the noise floor is constant across the memory array and independent of the number of cells and stored qubits~\cite{Teller2025a}.} Another path to increase the signal-to-noise ratios is to increase the efficiency of the memory array, e.g. by using longer crystals \cite{Hedges2010} or embedding the crystal into a spatially multimode cavity \cite{Wang2023,Duranti2024}. \added{We highlight that a longer crystal may be integrated without affecting the scalability of this spatial multiplexing approach.} While the memory efficiencies of the array are close to the values for single-cell experiments, the overall efficiency is limited by the fiber coupling to the detection setup after the demultiplexing AOD. Some modes have half the fiber coupling efficiency than others~\cite{SuppMat}, and we expect that by optimizing the optical components for this experimental setup, we will achieve a uniform fiber coupling efficiency as demonstrated with cold-atomic memory arrays~\cite{Pu2017}. With a uniform fiber coupling efficiency, we expect to achieve storage fidelities of path and time-bin qubits exceeding $90\;\%$ for the full array. In near future, the number of available cells can be increased from ten to hundreds by using two-dimensional AODs~\cite{Pu2017,Zhang2024}. \added{With each cell acting as an independent temporally-multimode quantum memory, storage in thousands of spatio-temporal modes may thus be within reach}. \added{While in the presented experiments the shortest possible storage time of $\SI{14}{\micro\second}$ is limited by the AFC storage time and the duration of the two control pulses, we emphasize that both can be significantly reduced to achieve a minimum storage time less than $\SI{5}{\micro\second}$, at the expense of reduced temporal multimodality.} \deleted{Lastly,} \replaced{T}{t}he \added{maximum} storage time of the qubits was limited to tens of microseconds due to the \added{spin inhomogeneous broadening}. Dynamical decoupling techniques offer a path to long storage times of qubits~\cite{Ortu2022a, Liu2025} and could be applied to the array with radio-frequency or optical pulses. \added{Lastly, the AFC protocol implemented in each memory cell introduces a latency between the control pulse requesting the stored qubit and the emission from the memory array.}

Our results provide a significant advance towards a RAQM based on rare-earth-doped quantum memory arrays.
Thanks to the intrinsic temporal multimodality of the quantum memory cells, our array can store quantum information encoded in $d$ dimensions of a time-bin qudit in a single cell~\cite{Holzapfel2023}, something which in other platforms would require mapping on $d$ individual cells~\cite{Li2020c}. Deployed in the future quantum internet~\cite{Kimble2008}, qudits may offer higher secret key rates and noise-resilience in quantum cryptography~\cite{Bechmann-Pasquinucci2000,Cozzolino2019} and may help to connect distant qudit processors~\cite{Ringbauer2022,Morvan2021,FernandezdeFuentes2024}. Furthermore, \added{with improved efficiencies} we envision applications of this memory array in the storage of photonic cluster states for linear quantum computing~\cite{Raussendorf2001}. Each memory cell would store a photonic cluster state, and through the here demonstrated collective or sequential read-out, joint measurements on both states could be implemented.

The data supporting the presented studies are available in Ref.~\cite{Zenodo2025}.

\section*{Acknowledgements}
\added{This work received financial support by:} Gordon and Betty Moore Foundation (GBMF7446 to H.~d.~R.); Agència de Gestió d'Ajuts Universitaris i de Recerca (AGAUR); Centres de Recerca de Catalunya; FUNDACIÓ Privada MIR-PUIG; Fundación Cellex; 
\added {Government of Spain, Ministerio de Ciencia e Innovación with funding from European Union NextGeneration funds (MCIN/AEI/10.13039/501100011033, PLEC2021-007669 QNetworks, PRTR-C17.I1); Agencia Estatal de Investigación (PID2019-106850RB-100, PID2023-147538OB-I00, Severo Ochoa CEX2019-000910-S)};
 European Union research and innovation program within the Flagship on Quantum Technologies through Horizon Europe project QIA-Phase 1 under grant agreement no. 101102140; European Union’s Horizon 2020 Research and Innovation Programme under the Marie Skłodowska-Curie grant agreement number 956419 (NanoGlass); Secretariat of Digital Policies of the Government of Catalonia - G.A. GOV/51/2022. S.G. acknowledges funding from ``La Caixa'' Foundation (ID 100010434, fellowship LCF/BQ/PR23/11980044); M.T. acknowledges funding from the European Union's Horizon 2022 research and innovation programme under the Marie Sklodowska-Curie grant agreement No 101103143 "Two-dimensionally multiplexed on-demand quantum memories" (2DMultiMems).
 
The published version of record of this manuscript is available open access in Ref.~\citep{Teller2025b}.

\bibliography{qpsa.bib}

\end{document}